\documentclass{article}
\usepackage{spconf,amsmath,graphicx}
\usepackage{booktabs}
\usepackage[breaklinks=true]{hyperref}

\title{TACKLING THE SCORE SHIFT IN CROSS-LINGUAL SPEAKER VERIFICATION BY EXPLOITING LANGUAGE INFORMATION}

\name{Jenthe Thienpondt, Brecht Desplanques, Kris Demuynck}
\address{IDLab, Department of Electronics and Information Systems, Ghent University - imec, Belgium}
\begin{document}
\ninept
\maketitle
%


\begin{abstract}
This paper contains a post-challenge performance analysis on cross-lingual speaker verification of the IDLab submission to the VoxCeleb Speaker Recognition Challenge 2021 (VoxSRC-21). We show that current speaker embedding extractors consistently underestimate speaker similarity in within-speaker cross-lingual trials. Consequently, the typical training and scoring protocols do not put enough emphasis on the compensation of intra-speaker language variability. We propose two techniques to increase cross-lingual speaker verification robustness. First, we enhance our previously proposed Large-Margin Fine-Tuning (LM-FT) training stage with a mini-batch sampling strategy which increases the amount of intra-speaker cross-lingual samples within the mini-batch. Second, we incorporate language information in the logistic regression calibration stage. We integrate quality metrics based on soft and hard decisions of a VoxLingua107 language identification model. The proposed techniques result in a 11.7\% relative improvement over the baseline model on the VoxSRC-21 test set and contributed to our third place finish in the corresponding challenge.

\end{abstract}
\section{Introduction}
\label{sec:intro}

The goal of speaker verification is to determine if two utterances are uttered by the same person. Currently, typical speaker verification systems use low-dimensional speaker embeddings extracted from speaker identification models based on Time Delay Neural Networks (TDNNs)~\cite{x_vectors, x_vector_wide, ecapa_tdnn} or ResNet~\cite{resnet, magneto, freq_paper} architectures. The advent of margin- and angular-based loss functions such as Additive Margin (AM)~\cite{am_softmax} and Additive Angular Margin (AAM)~\cite{arcface} softmax enables the use of cosine similarity between embeddings to score speaker similarity. These neural network based speaker identification models are trained on large datasets of labelled speech utterances to create robust speaker embeddings. A popular dataset is the development part of the VoxCeleb2~\cite{vox2} corpus, which contains over 1 million utterances from 5994 speakers.

Speaker verification systems should be robust against cross-lingual trial conditions and discriminate between speakers independently of the language spoken.
However, spoken language or dialect could be modelled as a speaker characterizing feature by the neural network. As a result, speaker verification systems are prone to underestimating the speaker similarity in positive (within-speaker) cross-lingual trials.
This effect is enhanced by the domination of speakers from the Anglosphere and limited intra-speaker linguistic variability in current popular speaker identification datasets.

The VoxCeleb Speaker Recognition Challenge 2021 (VoxSRC-21)~\cite{voxsrc_2021} aims to provide a challenging speaker verification test set with an emphasis on cross-lingual trials.
The competition rules allow to incorporate information from a pre-trained language classification model to improve robustness against cross-lingual conditions.
In this paper we analyze and further develop the cross-linguality compensation techniques we used in our VoxSRC-21 track 1 submission~\cite{idlab_voxsrc21_td}.

We propose two enhancements to increase intra-speaker cross-lingual robustness. Both techniques exploit information from a language classification model. First, we propose a cross-lingual fine-tuning stage to make the speaker embedding extractor more robust against varying phonetic content. Second, we introduce and analyse the addition of language information in our previously proposed quality-aware score calibration stage \cite{icassp_voxsrc20}.

The paper is organized as follows: Section~\ref{sec:baseline} describes the baseline speaker verification system.
Section~\ref{sec:cl_ft} and \ref{sec:calibration} outline our proposed cross-lingual fine-tuning stage and language-aware calibration system, respectively. Section~\ref{sec:setup} describes the experimental setup we use to validate our proposed enhancements. Subsequently, Section~\ref{sec:results} will discuss the results of the experiments. Finally, Section~\ref{sec:conclusion} will give some concluding remarks.


\section{Baseline System}
\label{sec:baseline}




We choose the best performing single system from our final submission on the VoxSRC-21 validation set~\cite{idlab_voxsrc21_td} as our baseline. The architecture of this fwSE-ResNet model is inspired by~\cite{magneto} and incorporates frequency-wise Squeeze-Excitation (fwSE) and frequency positional encodings~\cite{freq_paper}. The topology is defined in Table~\ref{tab:res}. Standard ResNet models are based on 2D convolutions, resulting in frequency- and time-equivariance of the model. However, speaker-specific speech patterns are expected to be different across lower and higher frequency regions. This makes the addition of frequency positional encodings in the network beneficial. The ResNet architecture is further enhanced to process speech by modifying the Squeeze-Excitation module to rescale activations frequency-wise instead of using the standard channel-wise rescaling. More information can be found in~\cite{freq_paper}.

Our baseline speaker verification system is fine-tuned using the Large-Margin Fine-Tuning (LM-FT) protocol~\cite{icassp_voxsrc20}. This secondary training stage increases the margin penalty of the AAM-softmax criterion to enforce greater inter-speaker distances and decrease the intra-speaker variability of the embeddings. The increased training difficulty caused by the higher margin configuration is compensated by taking longer fixed-length crops of the training utterances during fine-tuning.

Finally, the speaker verification trial scores are calibrated by the quality-aware score calibration backend described in~\cite{icassp_voxsrc20}. This calibration stage converts the raw trial scores to proper log-likelihood-ratios. This post-processing step also increases the speaker discriminatory ability of the system by compensating for varying quality conditions of the recordings in the trials.

\begin{table}[!h]
    \fontsize{8.5}{10}\selectfont
  \centering
  \begin{tabular}{lcr}
    \toprule
     Layer name & Structure & \multicolumn{1}{r}{Output} \\
      & & \multicolumn{1}{r}{$C$$\times$$F$$\times$$T$} \\
    \midrule
    \midrule
    log Mel-FBE & - & 1$\times$80$\times$T \\
    \midrule
    Conv2D & 3 $\times$ 3, stride=1 & 128$\times$80$\times$T \\ 
    \midrule
    ResBlock1 & $\begin{bmatrix} \textrm{pos. enc.}, 80\\ 3\times3, 128\\ 3\times3, 128\\\textrm{fwSE}, [128, 80] \end{bmatrix}\times$12, stride=1 & 128$\times$80$\times$T \\
    \midrule
    ResBlock2a & $\begin{bmatrix} \textrm{pos. enc.}, 80\\ 3\times3, 128\\ 3\times3, 128\\\textrm{fwSE}, [128, 40]\\ \overline{1\times1, 128}\end{bmatrix}\times$1, stride=2 & 128$\times$40$\times$T/2 \\
    \addlinespace[0.6ex]
    ResBlock2b & $\begin{bmatrix} \textrm{pos. enc.}, 40\\ 3\times3, 128\\ 3\times3, 128\\\textrm{fwSE}, [128, 40]\end{bmatrix}\times$15, stride=1 & 128$\times$40$\times$T/2 \\
    \midrule
    ResBlock3a & $\begin{bmatrix} \textrm{pos. enc.}, 40\\ 3\times3, 256\\ 3\times3, 256\\\textrm{fwSE}, [128, 20]\\ \overline{1\times1, 256}\end{bmatrix}\times$1, stride=2 & 256$\times$20$\times$T/4 \\
    \addlinespace[0.6ex]
    ResBlock3b & $\begin{bmatrix} \textrm{pos. enc.}, 20\\ 3\times3, 256\\ 3\times3, 256\\\textrm{fwSE}, [128, 20]\end{bmatrix}\times$11, stride=1 & 256$\times$20$\times$T/4 \\
    \midrule
    ResBlock4a & $\begin{bmatrix} \textrm{pos. enc.}, 20\\ 3\times3, 256\\ 3\times3, 256\\\textrm{fwSE}, [128, 10]\\ \overline{1\times1, 256}\end{bmatrix}\times$1, stride=2 & 256$\times$10$\times$T/8 \\
    \addlinespace[0.6ex]
    ResBlock4b & $\begin{bmatrix} \textrm{pos. enc.}, 10\\ 3\times3, 256\\ 3\times3, 256\\\textrm{fwSE}, [128, 10]\end{bmatrix}\times$2, stride=1 & 256$\times$10$\times$T/8 \\
    \midrule
    Flatten (C, F) & - & 2560$\times$T/8 \\
    CAS pooling  & - & 5120 \\
    \midrule
    Linear (emb.) & - & 256 \\
    AAM-softmax & & \#Speakers \\
    \bottomrule
  \end{tabular}
    \caption{The fwSE-ResNet architecture based on~\cite{magneto}  with frequency-wise Squeeze-Excitation and frequency positional encodings~\cite{freq_paper}. $C$, $F$ and $T$ are the channel, frequency and time dimensions, respectively. The pooling is realized by Channel-dependent Attentive Statistics (CAS)~\cite{ecapa_tdnn}. The 1 $\times$ 1 convolutions are used in the residual connections to match dimensions of the activation maps.}
    \label{tab:res}
\end{table}

\section{Cross-lingual fine-tuning}
\label{sec:cl_ft}

We want the speaker embeddings to be invariant to varying phonetic content and variation in spoken language. However, most speakers in the dataset will have a limited amount of spoken language variability. Subsequently, the model will likely interpret the spoken language or dialect as a speaker characterizing feature. We argue this could make the model underestimate speaker similarity in cross-lingual trials.

To mitigate this, we propose a cross-lingual fine-tuning stage. 
In this training stage we increase the intra-speaker language variability on the mini-batch level. 
Instead of sampling utterances randomly, we iteratively construct cross-lingual mini-batches. 

We combine this strategy with LM-FT and replace the hard sampling algorithm of LM-FT with cross-lingual sampling. First, the spoken language of an utterance is estimated using a language classification model. This enables the selection of cross-lingual utterance pairs. Subsequently, mini-batches are constructed by randomly iterating over all $N$ training speakers in our dataset. During an iteration, each mini-batch contains samples from $S$ speakers with each $U$ cross-lingual utterances. The cross-lingual utterances are selected in pairs of two, alleviating the need to have a large amount of mutually cross-lingual utterances for each training speaker in case $U>2$. We resort to random sampling when a speaker does not have any cross-lingual utterance pairs available. A single iteration continues until all training speakers are processed, after which the procedure is repeated.

\section{Language calibration}
\label{sec:calibration}

Score calibration backends in speaker verification systems convert speaker similarity scores to well-calibrated log-likelihood-ratios~\cite{bosaris}. Calibration based on logistic regression has recently proved to improve speaker verification performance for DNN-based systems by including Quality Metric Functions (QMFs)~\cite{icassp_voxsrc20, alenin21_interspeech, speakin}.

Quality-aware score calibration \cite{icassp_voxsrc20} learns a mapping $l$ from the speaker similarity score $s$ to produce a calibrated log-likelihood-ratio $l(s)$. The mapping is defined as $l(s) = w_{s}s + \textbf{w}_{q}^{T} \textbf{q} + b$ with $w_s$ and $\textbf{w}_{q}$ being learnable weights for the trial score $s$ and the quality features $\textbf{q}$, respectively. Since this mapping is non-monotonic, it can improve metrics with a fixed decision threshold such as EER and MinDCF.

\begin{figure}[h]
\begin{minipage}[h]{1.0\linewidth}
  \centering
  \centerline{\includegraphics[scale=0.28]{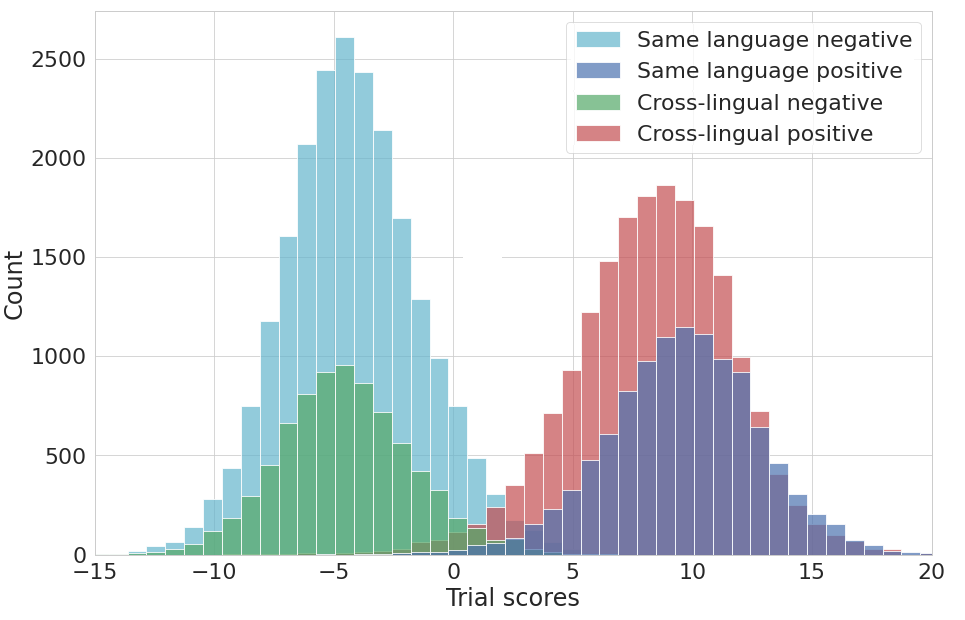}}
\end{minipage}
\caption{Histogram of the s-normalized fwSE-ResNet trial scores on the VoxSRC-21 validation set.}
\label{fig:score_shift}
\end{figure}

Figure~\ref{fig:score_shift} shows a histogram of s-normalized~\cite{s_norm, s_norm_2} trial scores of the baseline fwSE-ResNet system on the VoxSRC-21 validation set. Cross-lingual trials are defined according to language labels provided by the challenge organizers. The figure clearly shows that the baseline system underestimates speaker similarity under cross-lingual trial conditions. We propose to add language features based on hard or soft output decisions of a language classification model to allow the calibration backend to compensate for the score shift induced by cross-linguality. 
A range of potential language features are discussed in the subsections below.

\subsection{Binary cross-linguality indicator}
\label{sec:lang_binary}

We can use the classification output of the language classifier to determine the most probable spoken language of an utterance. Subsequently, we construct a binary feature indicating if the predicted language of the enrollment and the test side of a speaker verification trial are the same or not. 

\subsection{Similarity of predicted language class probabilities}
\label{sec:lang_js}

A cross-lingual binary feature has some limitations. First, the language classification model is prone to errors, especially on languages with a limited amount of training data~\cite{voxlingua}. In addition, a binary feature does not express any uncertainty on the language estimation of the model. Second, it only provides information with regards to the predicted spoken language, neglecting potential information about the likeliness of the utterance to other languages.


To mitigate these issues, we construct a language feature using the output probabilities $E$ and $T$ of the language classification model of the enrollment and test side utterance of the speaker verification trial, respectively. 
In case of AAM-softmax trained models, we obtain the probabilities by scaling the output cosine distances of the language classifier by the proper AAM scale factor, followed by a softmax operation.

We want our language calibration feature to be side-independent of the trial, making most divergence-based metrics of the output probabilities unsuitable.
We propose the Jensen-Shannon distance between both language classification probabilities as a calibration feature as it obeys the symmetry requirement. The Jensen-Shannon distance can be regarded as a symmetrical and smoother version of the Kullback-Leibler~\cite{KL} divergence. Given both language classification output distributions, the Jensen-Shannon distance is defined as:
\begin{equation}
JS(E\|T) = \sqrt{\frac{D_{KL}(E \| M)+D_{KL}(T \| M)}{2}}
\end{equation}
with $M$ equal to $\frac{E+T}{2}$ and $D_{KL}$ indicating the Kullback-Leibler divergence.




\subsection{Similarity of language embeddings}
\label{sec:lang_emb}

Language features based on the classification probabilities directly rely on the confusion of the classifier between the language classes to model similarities between languages. This could negatively impact the ability to express intra-language variability (e.g. dialects of the same language) and language information from unseen classes. However, the final linear layers of the language classification neural network potentially contain a more general and expressive representation of the spoken language.
When the language classification model is trained using an angular-based loss function, such as the AAM-softmax, low-dimensional language embeddings can be extracted from the final linear projection layer. The spoken language of the utterances can be directly compared using the cosine distance between the extracted language embeddings. Scoring language embeddings should also generalize better when encountering new languages not seen during the training of the language classifier. Subsequently, we propose the cosine distance of the language embeddings of the enrollment and test side of the trial as a calibration feature.

\begin{table*}
  \centering
  \begin{tabular}{clcccccccc}
    \toprule
    
    \multicolumn{1}{c}{} &
    \multicolumn{1}{c}{} &
    \multicolumn{2}{c}{\textbf{Cross-lingual}} &
    \multicolumn{6}{c}{\textbf{Standard Benchmarks}} \\
    \cmidrule(lr){3-4} \cmidrule(lr){5-10}
    
    \multicolumn{1}{c}{} &
    \multicolumn{1}{l}{\textbf{System Configuration}} &
    \multicolumn{2}{c}{\textbf{VoxSRC-21 Val}} &
    \multicolumn{2}{c}{\textbf{VoxCeleb1-O}} &
    \multicolumn{2}{c}{\textbf{VoxCeleb1-E}} & 
    \multicolumn{2}{c}{\textbf{VoxCeleb1-H}} \\

    \cmidrule(lr){3-4} \cmidrule(lr){5-6} \cmidrule(lr){7-8} \cmidrule(lr){9-10}
    \multicolumn{2}{c}{\textbf{}} & 
    \multicolumn{1}{c}{\textbf{EER(\%)}} & \multicolumn{1}{c}{\textbf{MinDCF}} &
    \multicolumn{1}{c}{\textbf{EER(\%)}} & \multicolumn{1}{c}{\textbf{MinDCF}} &
    \multicolumn{1}{c}{\textbf{EER(\%)}} & \multicolumn{1}{c}{\textbf{MinDCF}} &
    \multicolumn{1}{c}{\textbf{EER(\%)}} & \multicolumn{1}{c}{\textbf{MinDCF}}\\
    
    \midrule
     & fwSE-ResNet & 2.82 & 0.1538 & 0.64 & 0.0489 & 0.84 & 0.0925 & 1.51 & 0.1471 \\
     & fwSE-ResNet + LM-FT & 2.41 & 0.1343 &  0.55 & 0.0383 &  0.76 & 0.0824 & 1.35 & 0.1300 \\
     & fwSE-ResNet + CL LM-FT & 2.25 & 0.1234 & 0.58 & 0.0375 & 0.74 & 0.0800 & 1.30 & 0.1228 \\
     \midrule
     & + log duration QMF & 2.11 & 0.1143 & \textbf{0.50} & \textbf{0.0377} & \textbf{0.71} & \textbf{0.0777} & \textbf{1.26} & \textbf{0.1204} \\
    \midrule
    \midrule
     & ++ binary QMF (\ref{sec:lang_binary}) & 1.84 & 0.1038 & 0.58 & 0.0639 & 0.78 & 0.0843 & 1.42 & 0.1436 \\ 
     & ++ Jensen-Shannon QMF (\ref{sec:lang_js}) & 1.67 & 0.0899 & 0.59 & 0.0586 & 0.77 & 0.0837 & 1.38 & 0.1366 \\ 
     & ++ cosine distance QMF (\ref{sec:lang_emb}) & \textbf{1.63} & \textbf{0.0827} & 0.55 & 0.0539 & 0.74 & 0.0794 & 1.30 & 0.1274 \\
    \bottomrule
  \end{tabular}
    \caption{Analysis of cross-lingual fine-tuning and calibration with language information of the fwSE-ResNet system.}
  \label{tab:calibration_results}
\end{table*}

\section{Experimental setup}
\label{sec:setup}

To analyse the performance impact of the proposed cross-lingual fine-tuning stage and the integration of the language calibration features, we apply our proposed enhancements on the baseline speaker verification system described in Section~\ref{sec:baseline}.

\subsection{Training configuration}
\label{sec:training_conf}

The baseline speaker embedding extractor is trained on the development part of VoxCeleb2. During training, we take random crops of two seconds of each utterance and apply a random augmentation using the MUSAN corpus~\cite{musan} (babble, music, noise) and the RIR~\cite{rirs} dataset (reverb) to prevent overfitting. The input features consist of 80-dimensional log Mel-filterbank energies (Mel-FBE) with a window length of 25 ms and a frame shift of 10 ms. To further enhance robustness, we apply SpecAugment \cite{specaugment} which randomly masks 0 to 10 frequency bands and 0 to 5 frames in the time-domain. Subsequently, all filterbank energies are mean normalized per utterance. A mini-batch size of 128 is used during training.

The model is trained using the Adam optimizer \cite{adam} with a cyclical learning rate \cite{clr} using the \textit{triangular2} policy with the minimum and maximum learning rate varying between 1e-8 and 1e-3, respectively. The cycle length is set to 130k. A weight decay of 2e-5 is used to regularize the model during training. The system is trained for one cycle with the AAM-softmax loss function using a margin and scale value of 0.2 and 30, respectively.

\subsection{Cross-lingual large-margin fine-tuning}

After the initial training phase, we apply cross-lingual LM-FT (CL LM-FT) on the model to create more discriminative speaker embeddings. In this stage, the crop size is extended to four seconds with a simultaneous AAM-softmax margin increase to 0.4. We use these settings as opposed to the originally proposed configuration in~\cite{icassp_voxsrc20} for computational reasons. Additionally, we change the random sampling of training utterances to the cross-lingual sampling strategy described in Section~\ref{sec:cl_ft}. We keep the initial batch size of 128 and vary the ratio of speakers $S$ and cross-lingual utterances $U$. We do not change the augmentation strategy.

\subsection{Quality and language aware calibration}

After the fine-tuning stage, speaker verification trial sores are normalized using adaptive s-normalization \cite{s_norm, s_norm_2} with an imposter cohort size of 400 speakers. Subsequently, we apply quality-aware score calibration using the log duration QMF~\cite{idlab_voxsrc21_td}.

We apply our proposed language calibration by adding the language features from Section~\ref{sec:calibration} to the calibration backend. We evaluate three types of language features based on the similarity between either output language predictions, output language probabilities or language embeddings. The language information is extracted from each utterance by an ECAPA-TDNN~\cite{ecapa_tdnn} language classifier\footnote{\url{https://huggingface.co/TalTechNLP/voxlingua107-epaca-tdnn}} pre-trained on VoxLingua107~\cite{voxlingua} using the AAM-softmax loss.

The calibration backend is trained on a custom VoxCeleb2 subset with half of the utterances cropped between 2 and 4 seconds. We initially select 100k trials and balance the amount of positive and negative trials. Half of the trials are cross-lingual. We discard 20\% of both positive and negative trials with the least and greatest cosine distance between the trial language embeddings, respectively. We apply this selection to compensate for overfitting induced by the fact that VoxCeleb2 is also the training dataset of the speaker embedding extractor. We only generate within-gender trials and did not include positive trials with utterances originating from the same video. 
\subsection{Evaluation protocol}

We evaluate the baseline system and proposed enhancements on the VoxCeleb test sets \cite{vox1, vox2} and report the EER and MinDCF metric using a $P_{target}$ value of $10^{-2}$ with $C_{FA}$ and $C_{Miss}$ equal to 1. To analyse the proposed techniques on challenging cross-lingual data, we also evaluate the systems on the VoxSRC-21 validation and test set using the challenge MinDCF metric with a $P_{target}$ value of $0.05$.

\section{Results}
\label{sec:results}
Table~\ref{tab:cl_ft} shows the performance impact of the proposed cross-lingual fine-tuning on the VoxSRC-21 validation set. The ratio of speakers and cross-lingual utterances ($S/U$) within a mini-batch during cross-lingual sampling is indicated between brackets. Standard LM-FT with random samples results in a relative performance improvement over the baseline model of 14.5\% and 12.7\% in EER and MinDCF, respectively. The most optimal cross-lingual sampling strategy uses 64 speakers with each two cross-lingual utterances in the mini-batch and improves performance further with a relative improvement of 6.6\% in EER and 8.1\% in MinDCF. As shown in Table~\ref{tab:cl_ft}, selecting more than one cross-lingual utterance pair per speaker on the mini-batch level is less effective. This is probably caused by the fact that we keep the mini-batch size constant due to computational limitations. Therefore, we have to reduce $S$ when we increase $U$. Moreover, the amount of intra-speaker cross-lingual utterances is limited in the training set and setting $U>2$ might not be optimal for every training speaker.

\begin{table}[h]
  \centering
  \begin{tabular}{lccc}
    \toprule
     \textbf{Method} & \textbf{Sampling} & \multicolumn{1}{c}{\textbf{EER(\%)}} & \multicolumn{1}{c}{\textbf{MinDCF}} \\
    \midrule
     baseline & random & 2.82 & 0.1538 \\ 
    \midrule
    LM-FT & random & 2.41 & 0.1343 \\ 
    \midrule
    LM-FT & cross-lingual (16/8) & 2.32 & 0.1241 \\ 
    LM-FT & cross-lingual (32/4) & 2.26 & 0.1237 \\ 
    LM-FT & cross-lingual (64/2) & \textbf{2.25} & \textbf{0.1234} \\ 
    \bottomrule
  \end{tabular}
    \caption{Evaluation of different configurations of cross-lingual fine-tuning on the VoxSRC-21 validation set.}
    \label{tab:cl_ft}
\end{table}

In Table~\ref{tab:calibration_results} we analyse the impact of the proposed language features in the calibration stage. In most cases, incorporating language features in the calibration backend results in a minimal performance degradation on the standard VoxCeleb test sets. Mistakes made by the language classifier cannot be sufficiently compensated by better cross-lingual performance due to the limited amount of cross-lingual trials in the standard benchmark datasets. However, we see a significant reduction of the cross-lingual score shift on the VoxSRC-21 validation set. Both the probability- and embedding-based features outperform the binary cross-lingual measure, showing that the system can effectively exploit the additional language information in the soft decision features. The cosine distance between the language embeddings performs the best with a relative improvement of 22.6\% and 27.6\% of the EER and MinDCF metric on the VoxSRC-21 validation set, respectively.
These results indicate that the addition of language calibration features is currently to be considered as a performance trade-off that should be acceptable in most use-cases.

\begin{table}[h]
  \centering
  \begin{tabular}{lcc}
    \toprule
     \textbf{Systems} & \multicolumn{1}{c}{\textbf{EER(\%)}} & \multicolumn{1}{c}{\textbf{MinDCF}} \\
    \midrule
    baseline + LM-FT + QMF & 2.78 & 0.1690 \\
    baseline + CL LM-FT + lang QMF & \textbf{2.72} & \textbf{0.1492} \\
    \bottomrule
  \end{tabular}
    \caption{Performance analysis of the proposed cross-lingual fine-tuning and language calibration on the VoxSRC-21 test set.}
    \label{tab:performance_test_set}
\end{table}

Finally, we evaluate the cross-lingual fine-tuning and language calibration performance impact on the VoxSRC-21 test set. Table~\ref{tab:performance_test_set} compares the LM-FT strategy with random sampling and quality-aware score calibration without language features against LM-FT with cross-lingual sampling and score calibration with language embeddings. Incorporating language information results in a relative improvement on the VoxSRC-21 test set of 11.7\% on the MinDCF challenge metric. This improvement is less significant than the observed performance increases on the VoxSRC-21 validation set.
We suspect this is mainly due to the significantly smaller crops ($<$ 4 seconds) in the test set, which could deteriorate the language information extracted by the VoxLingua language classifier. 

\section{Conclusion}
\label{sec:conclusion}

We proposed two enhancements in speaker verification to increase the robustness against cross-lingual trials. First, we introduced cross-lingual data sampling during fine-tuning of the embedding extractor. Second, we incorporated language information in the calibration backend to compensate for score shifts induced by cross-lingual conditions. By combining both strategies, we improved the baseline model relatively with 11.7\% on the main MinDCF metric on the challenging cross-lingual VoxSRC-21 test set.




\vfill\pagebreak

\bibliographystyle{IEEEbib}
\bibliography{refs}

\end{document}